\documentclass{article}
\usepackage{amsmath}
\usepackage{amsthm}
\usepackage{longtable}
\usepackage{multirow}
\usepackage{url}
\usepackage{booktabs}

\theoremstyle{definition} \newtheorem{definition}{Definition}
\theoremstyle{definition} \newtheorem{exampleenv}{Example}

\newcommand{\Exact}{\textsf{Exact}}
\newcommand{\SD}{\textsf{SD}}
\newcommand{\GKK}{\textsf{GKK}}
\newcommand{\GK}{\textsf{GK}}
\newcommand{\SG}{\textsf{SG}}

\newcommand{\dist}{\text{dist}}

\newcommand{\keywords}[1]{\par\addvspace\baselineskip
\noindent \textbf{Keywords}\enspace\ignorespaces#1}

\begin{document}

\title{Generalized Traveling Salesman Problem Reduction Algorithms}

\date{}

\author{Gregory Gutin\thanks{
Department of Computer Science,
Royal Holloway University of London,
Egham, Surrey TW20 0EX, UK,
G.Gutin@rhul.ac.uk}
\and Daniel Karapetyan\thanks{
Department of Computer Science,
Royal Holloway University of London,
Egham, Surrey TW20 0EX, UK,
Daniel.Karapetyan@gmail.com}}


\maketitle

\begin{abstract}
The generalized traveling salesman problem (GTSP) is an extension of the well-known traveling salesman problem. In GTSP, we are given a partition of cities into groups and we are required to find a minimum length tour that includes exactly one city from each group.  The aim of this paper is to present a problem reduction algorithm that deletes redundant vertices and edges, preserving the optimal solution.  The algorithm's running time is $O(N^3)$ in the worst case, but it is significantly faster in practice. The algorithm has reduced the problem size by 15--20\% on average in our experiments and this has decreased the solution time by 10--60\% for each of the considered solvers.
\end{abstract}

\section{Introduction}

The \emph{generalized traveling salesman problem} (GTSP) is defined
as follows. We are given a weighted complete undirected graph $G$ on
$N$ vertices and a partition $V = V_1 \cup V_2 \cup \ldots \cup V_M$
of its vertices; the subsets $V_i$ are called {\em clusters}. The
objective is to find a minimum weight cycle containing exactly one
vertex from each cluster. There are many publications on GTSP (see,
e.g., the surveys~\cite{fischetti2002,gutin2003} and the references
therein). The problem has many applications, see,
e.g.,~\cite{Transformation, Applications}. It is NP-hard, since the
\emph{traveling salesman problem} (TSP) is its special case (when
$|V_i| = 1$ for each $i$). The weight of an edge $xy$ of $G$ is
denoted $\dist(x,y)$ and will be often called the {\em distance}
between $x$ and $y$.

Various approaches to GTSP have been studied. There are exact
algorithms such as branch-and-bound and branch-and-cut described
in~\cite{BranchAndCut}. Another approach uses the fact that GTSP can
be converted to an equivalent TSP with the same number of
vertices~\cite{Transformation, TransformationLaporte,
TransformationLien, TransformationNoon} and then can be solved with
some efficient TSP solver such as \texttt{Concorde}~\cite{Concorde}.
Heuristic GTSP algorithms have also been invistigated, see,
e.g.,~\cite{GutinKarapetyanGA, GutinKarapetyanGA2009, HuangGA, GoldenGA, RandomKeyGA, FatihGA, ZhaoGA}.

Different preprocessing procedures are often used for hard problems to reduce the computation time. There are examples of such approaches in integer and linear programming (e.g.,~\cite{LPPreporcessing, IPPreporcessing}) as well as for the Vehicle Routing Problem~\cite{RoutingPreporcessing}. In some cases preprocessing plays the key role in an algorithm (e.g.,~\cite{LPPresolve}). We introduce preprocessing procedure for GTSP.  A feature of GTSP is that not every vertex of a problem should be visited and, thus, GTSP may contain vertices that a priori are not included in the optimal solution and may be removed.  We have a similar situation with edges.

The experimental results show that almost each GTSP instance tested in the literature can be reduced by the presented procedure at a very low cost and that this reduction is almost always beneficial for the GTSP solvers.

\section{Vertex Reduction}
\label{sec:VertexReduction}

Since GTSP solution covers only $M$ vertices, up to $N - M$ vertices may be reduced without a change of the optimal solution.  We present an approach to detect some of the redundant vertices in a reasonable time.

\begin{definition} \label{definition:excess}
Let $C$ be a cluster, $|C| > 1$. We say that a vertex $r \in C$ is \emph{redundant} if for each pair $x, y$ of vertices from distinct clusters different from $C$, there exists a vertex $s \in C \setminus \{ r \}$ such that $\dist(x, s) + \dist(s, y) \leq \dist(x, r) + \dist(r, y)$.
\end{definition}

In other words, if for each path $x r y$ there exists another path $xsy$, $s \in C \setminus \{ r \}$, with the same or smaller weight, vertex $r$ can be removed.  Testing this condition for every vertex will take approximately $O(N^3 \cdot \overline{|V|})$, where $\overline{|V|} = N / M$ is the average cluster size.  In the symmetric case of the problem there is an efficient heuristic that usually allows to reduce the preprocessing time significantly.

Let us take two distinct vertices $r$ and $s$ in some cluster $C$.  We can calculate the differences between the distances to $r$ and $s$ from each vertex $x \notin C$ ($\Delta_{x}^{r, s} = \dist(x, r) - \dist(x, s)$) and save this information to a \emph{Differences Table} such as Table~\ref{tab:differences}. Notice that in
Table~\ref{tab:differences} we assume that clusters 1 and 2 have three vertices each and cluster 3 has two vertices, $r$ belongs to the first cluster and it is the first vertex in the cluster, i.e., vertex $s$ can be only the second and the third vertices of cluster 1.

\begin{table}[ht]
	\centering
    \caption{Differences Table example.}
    \label{tab:differences}
\begin{tabular}{c l r r r l r r l r}
\toprule{}
$s \backslash x$ & \hspace{1em} & cl.2 v.1 & cl.2 v.2 & cl.2 v.3 & \hspace{1em} & cl.3 v.1 & cl.3 v.2 & \hspace{1em} & Negative \# \\
\cmidrule(){1-10}
v.2  &&  $2$   &  $0$   &  $-1$  &&  $-3$  &  $4$   &&   $2$ \\
v.3  &&  $-1$  &  $-2$  &  $-1$  &&  $1$   &  $2$   &&   $3$ \\
\cmidrule(){1-10}
max  &&  $2$   &  $0$   &  $-1$  &&  $1$    &  $4$   &&       \\
\cmidrule(){3-5}  \cmidrule(){7-8}
&&  \multicolumn{3}{r}{$\min\{2, 0, -1\} = -1$} &&  \multicolumn{2}{r}{$\min\{1, 4\} = 1$}    &  \\
\bottomrule{}
\end{tabular}
\end{table}

Observe that a vertex $r$ is redundant if there is no pair of vertices from different clusters such that the sum of differences $\Delta$ (see above) for these vertices is negative for every $s$, i.e., $r$ is redundant if for every $x$ and $y$ there exists $s \in C \setminus \{ r \}$ such that $\Delta_{x}^{r, s} + \Delta_{y}^{r, s} \geq 0$, where $x$ and $y$ belong to distinct clusters. That is due to
\begin{align*}
\Delta_{x}^{r, s} &+ \Delta_{y}^{r, s} \\
   &= \dist(x, r) - \dist(x, s) + \dist(y, r) - \dist(y, s) \\
   &= \dist(x, r) + \dist(r, y) - \big(\dist(x, s) + \dist(s, y)\big)
\end{align*}

Therefore we need to check every pair of columns $(col_1, col_2)$ (except the pairs of columns corresponding to the same clusters) in the Differences Table $T_{row, col}$.  If $T_{1, col_1} + T_{1, col_2} < 0$, we check the second row ($T_{2, col_1} + T_{2, col_2}$).  If the result is still negative, we check the third row, etc.  If all the rows are checked and each time we obtain a negative sum, the vertex $r$ cannot be removed and the rest of the procedure may be skipped.

\begin{exampleenv}
In the example above (Table~\ref{tab:differences}) it is necessary to perform up to $6$ tests provided in Table~\ref{tab:pairs}.

\begin{table}[ht]
	\centering
   \caption{Vertices pairs for the example.}
   \label{tab:pairs}
\begin{tabular}{c @{\qquad} r @{\qquad} r}
	\toprule{}
    Pair                        & Sum for $s$ = v.2 & Sum for $s$ = v.3 \\
    \cmidrule{1-3}
    cl.2 v.1---cl.3 v.1 & $-1$              & $0$               \\
    cl.2 v.1---cl.3 v.2 & $6$               & $-1$               \\
    cl.2 v.2---cl.3 v.1 & $-3$              & $-1$              \\
    cl.2 v.2---cl.3 v.2 & $4$               & $0$               \\
    cl.2 v.3---cl.3 v.1 & $-4$              & $0$               \\
    cl.2 v.3---cl.3 v.2 & $3$               & $1$               \\
	\bottomrule{}
\end{tabular}
\end{table}

The only test that does not allow us to declare the vertex $r$ redundant is in the row 3 of the Table \ref{tab:pairs} (cl.2 v.2---cl.3 v.1) as both sums
(for $s = \text{v}.2$ and for $s = \text{v}.3$) are negative. (Certainly, there is no need to calculate the sum for $s$ = v.3 in rows 2, 4, and 6 in the example above, and the calculations may be stopped after the row 3.)

\end{exampleenv}

Removing redundant vertex may cause a previously irredundant
vertex to become redundant.  Thus, it is useful to check redundancy of vertices in cyclic order until we see that, in the last cycle, no vertices are found to be redundant. However, in the worst case, that would lead to $\Theta(N^2)$ redundancy tests.  (Recall that $N$ is the total number of vertices in GTSP.)  Our computational experience has shown that almost all redundant vertices will be found even if we restrict ourselves to testing each vertex of GTSP at most twice.  Thus, we assume in the rest of the paper that each vertex is tested at most twice for redundancy.

\subsection{Acceleration Heuristic}

In some cases it is possible to determine faster that a vertex $r$
is not redundant. If
$$
\min_{x \notin Z} \max_{s \in C} \Delta_{x}^{r, s} + \min_{x \in Z} \max_{s \in C} \Delta_{x}^{r, s} < 0
$$
for some cluster $Z$, then $r$ cannot be reduced. This condition
means that there exist two columns in the Differences Table
corresponding to distinct clusters and the sum of these columns
maxima is negative. This ensures that the sum for every row of these
columns is also negative.

We can use an equivalent condition:
$$
\min_{x \in \bigcup_{j < i} V_j} \max_{s \in C} \Delta_{x}^{r, s}
    + \min_{x \in V_i} \max_{s \in C} \Delta_{x}^{r, s} < 0
$$
This condition can be tested during the Difference Table generation. For each column we calculate the maximum value:
$$
vertexmax(x) = \max_{s \in C} \Delta_{x}^{r, s}
$$
Also for each cluster $Z$, we have
$$
clustermin(Z) = \min_{x \in Z} vertexmax(x)
$$
We define $totalmin(i)=\min_{j<i} clustermin(V_j)$; if $ totalmin(i)
+ clustermin(V_i) < 0 $ for some $i$, we can conclude that vertex
$r$ is not redundant.

In the example above, the heuristic performs just one check for
$V_2$ and $V_3$. We have $totalmin(3) = clustermin(V_2) = -1$ and
$clustermin(V_3) = 1$ and $-1 + 1 \geq 0$ so the acceleration
heuristic does not reduce our computations in this case.

Another way to make the redundancy test faster is to order the rows
of the Differences Table such that the row with the minimal number
of negative values would be the first one. Notice that, if this row
contains no negative values, it is obvious that $r$ is redundant.

\subsection{Algorithm Complexity}

Let $K_\text{min}$ and $K_\text{max}$ be the minimum and the maximum
number of tests (of vertices) for redundancy. Observe that
$K_\text{min} = N$, since we will perform only $N$ tests if no
vertex is detected to be redundant. Since we have assumed that no
vertex is tested more than twice for redundancy, $K_\text{max} = 2 N
- 1$.

Now consider how many operations are required for each redundancy test (with a fixed vertex $r$). The test requires table generation and table processing.  Due to the acceleration heuristic, table generation can be aborted already after processing of two clusters.  Thus, in the best case it takes $E_\text{min} = (|C| - 1) (|X| + |Y|)$~operations where $r \in C$, and $X$ and $Y$ are some other clusters. The average size of a cluster can be estimated as $N / M$ (recall that $M$ is the number of clusters). Therefore, in the best case each redundancy test requires approximately
$$
E_\text{min}(N')
    \approx \left(\frac{N'}{M} - 1\right) \left(2 \cdot \frac{N'}{M} \right)
    \approx 2 \cdot \left( \frac{N'}{M} \right)^2
$$
operations, where $N'$ is the current number of vertices in the problem.

In the worst case both the table generation and the further table inspection will be completed normally. Table generation will take $(|C| - 1)(N' - |C|)$ operations. Table inspection takes about $\left(N' - |C| - \overline{|V|} \right)^2 \left(|C| - 1 \right) / 2$ operations in the worst case, where $\overline{|V|}$ is the average cluster size. Thus, we have the following number of operations per test in the worst case:
\begin{align*}
E_{\max}(N')
    &\approx \left(|C| - 1\right) \left(N' - |C| \right)
        + \frac{\left(N' - |C| - \overline{|V|} \right)^2 \left(|C| - 1\right)}{2}  \\
    &\approx |C| \cdot N' + \frac{(N')^2 \cdot |C|}{2}
        \approx \frac{(N')^3}{2 M}.
\end{align*}

The total number of operations in the worst case is
$$
K_{\max} \cdot  E_{\max}(N)
     \approx 2 \cdot N \cdot \frac{N^3}{2 M}
     = \frac{N^4}{M}.
$$

The total operation number in the best case is
$$
K_{\min} \cdot E_{\min}(N)
     \approx N \cdot 2 \cdot \left( \frac{N}{M} \right)^2
     = 2 \cdot \frac{N^3}{M^2}.
$$

Since usually $M = \Theta(N)$, the algorithm complexity changes from $O(N)$ to $O(N^3)$. The experimental algorithm complexity is $\Theta(N^{2.4})$ (see Section~\ref{PureResults}).

\section{Edge Reduction}
\label{sec:EdgeReduction}

\begin{definition} \label{definition:excessedge}
Let $u$, $v$ be a pair of vertices from distinct clusters $U$ and $C$ respectively. Then the edge $uv$ is \emph{redundant} if for each vertex $x \in V \setminus U \setminus C$ there exists $v' \in C \setminus \{ v \}$ such that $\dist(u, v') + \dist(v', x) \leq \dist(u, v) + \dist(v, x)$.
\end{definition}

Testing this condition for every edge will work for both symmetric and asymmetric cases and will take approximately $O(N^3 \cdot \overline{|V|})$, where $\overline{|V|}$ is the average cluster size.  We introduce an algorithm for edge reduction for the symmetric case of the problem; it proceeds as follows.  Given a vertex $v \in C$, where $|C| > 1$, we detect redundant edges incident with $v$ using the following procedure:
\begin{enumerate}
    \item Select an arbitrary vertex $v'' \in C \setminus \{ v \}$.

    \item Set $P_x = \Delta_{x}^{v, v''}$ for each vertex $x \in V \setminus C$ (recall that $\Delta_{x}^{r, s} = \dist(x, r) - \dist(x, s)$).

    \item Sort array $P$ in non-decreasing order.

    \item For each cluster $U \neq C$ and for each vertex $u \in U$ do the following:
        \begin{enumerate}
            \item $\delta = \Delta_u^{v, v''}$
            \item \label{EdgeRedundantCheck} For each item $\Delta_{x}^{v, v''}$ of the array $P$ such that $\Delta_x^{v, v''} + \delta < 0$ check the following: if $x \notin U$ and $\Delta_x^{v, v'} + \Delta_{u}^{v, v'} < 0$ for every $v' \in C \setminus \{ v, v'' \}$, the edge $uv$ is not redundant, continue with the next $u$.
            \item Edge $uv$ is redundant, set $\dist(u, v) = \infty$.
        \end{enumerate}
\end{enumerate}

To prove that the above edge reduction algorithm works correctly, let fix some edge $uv$, $u \in U$, $v \in C$, $U \neq C$. The algorithm declares this edge redundant if the following condition holds for each $x \notin C$ (see \ref{EdgeRedundantCheck}):
\begin{align*} \label{EdgeRedundantCondition}
    \Delta_{x}^{v, v''} + \Delta_{u}^{v, v''} \geq 0 & \qquad \text{or}\\
    \Delta_{x}^{v, v'} + \Delta_{u}^{v, v'} \geq 0 & \qquad \text{for some } v' \in C \setminus \{ v, v'' \}
\end{align*}
This condition is equivalent to
$$
    \Delta_{x}^{v, v'} + \Delta_{u}^{v, v'} \geq 0 \qquad \text{for some } v' \in C \setminus \{ v \}
$$

So the algorithm declares the edge $uv$ redundant if for each $x \in V \setminus C \setminus U$ there exists $v' \in C \setminus \{ v \}$ such that $\Delta_x^{v, v'} + \Delta_u^{v, v'} \geq 0$.
\begin{align*}
    & \dist(x, v) - \dist(x, v') + \dist(u, v) - \dist(u, v') \geq 0 \qquad \text{and} \\
    & \dist(u, v) + \dist(v, x) \geq \dist(u, v') + \dist(v', x). \\
\end{align*}

Let us evaluate the algorithm's complexity. The edge reduction algorithm performs the following steps for every cluster $C$, $|C| > 1$ for each $v \in C$:
\begin{itemize}
    \item Array $P$ generation. This takes $\Theta(N)$ operation.
    \item Array $P$ sorting. This takes $\Theta(N \log_{2}{N})$ operations.
    \item Edges $uv$ testing. Each test takes $O(1)$ to $O(N \cdot |C|$) operations and $\Theta(N)$ tests are performed.
\end{itemize}

Thus the complexity of the entire algorithm is $\Theta (N^2 \log_{2}{N})$ in the best case, and $\Theta (N^3 \cdot |C|)$ in the worst case.

As usually $|C| = \Theta(N)$, we may say that this algorithm's complexity varies from $\Theta (N^2 \log_{2}{N})$ to $\Theta (N^3)$. The experimental algorithm complexity is $\Theta(N^{2.6})$ (see Section \ref{PureResults}).

After the search for redundant edges has been completed, the edge reduction algorithm finds redundant vertices using the following observation: if after the edge reduction procedure some vertex has finite distance edges to at most one cluster, then this vertex can be declared redundant.

This reduction takes $O(N^2)$ operations.

\section{Experiments}

We tested the reduction algorithms on the standard GTSP instances (see, e.g.,~\cite{Transformation, GoldenGA, RandomKeyGA, FatihGA}) which were generated from some \texttt{TSPLIB}~\cite{TSPLIB} instances by applying the clustering procedure of Fischetti, Salazar and Toth~\cite{BranchAndCut}. The algorithms were implemented in C++ and tested on a computer with AMD Atlon 64 X2 Core Dual processor ($3$ GHz frequency).

We have tested three reduction algorithms: the Vertex Reduction Algorithm (see Section \ref{sec:VertexReduction}), the Edge Reduction Algorithm (see Section \ref{sec:EdgeReduction}), and the Combined Algorithm witch first applies the Vertex Reduction Algorithm and then the Edge Reduction Algorithm.

\subsection{Experimental Results} \label{PureResults}

Each test was repeated ten times. The columns of the table are as follows:
\begin{itemize}
    \item \emph{Instance} is the instance name. The prefix number is the number of clusters of the instance; the suffix number is the number of vertices (before any preprocessing).
    \item $R_v$ is the number of vertices detected as redundant.
    \item $R_e$ is the number of edges detected as redundant. For the Combined Algorithm $R_e$ shows the number of redundant edges in the already reduced by the Vertex Reduction Algorithm problem.
    \item $T$ is the preprocessing time in seconds.
\end{itemize}

\begin{longtable}{l c r r r c r r r c r r r }
\caption{Test results of the Reduction Algorithms.} \label{tab:pure} \\
\toprule{}

    && \multicolumn{3}{c}{Vertex reduction}
    && \multicolumn{3}{c}{Edge reduction}
    && \multicolumn{3}{c}{Combined reduction} \\

\cmidrule{3-5}  \cmidrule{7-9}  \cmidrule{11-13}

Instance
    && $R_v$		& $R_v$, \%	& $T$
    && $R_e$, \%	& $R_v$		& $T$
    && $R_v$, \%,	& $R_e$, \%	& $T$ \\
\cmidrule{1-13}
\endfirsthead

    && \multicolumn{3}{c}{Vertex reduction}
    && \multicolumn{3}{c}{Edge reduction}
    && \multicolumn{3}{c}{Combined reduction} \\

\cmidrule{3-5}  \cmidrule{7-9}  \cmidrule{11-13}

Instance
    && $R_v$		& $R_v$, \%	& $T$
    && $R_e$, \%	& $R_v$		& $T$
    && $R_v$, \%,	& $R_e$, \%	& $T$ \\
\cmidrule{1-13}
\endhead

4ulysses16	&&	9	&	56.3	&	0.0	&&	62.0	&	4	&	0.0	&&	56.3	&	23.5	&	0.0	\\
4gr17     	&&	11	&	64.7	&	0.0	&&	35.8	&	3	&	0.0	&&	64.7	&	23.0	&	0.0	\\
5gr21     	&&	8	&	38.1	&	0.0	&&	48.7	&	3	&	0.0	&&	38.1	&	45.0	&	0.0	\\
5ulysses22	&&	11	&	50.0	&	0.0	&&	44.3	&	2	&	0.0	&&	50.0	&	39.5	&	0.0	\\
5gr24     	&&	13	&	54.2	&	0.0	&&	33.1	&	3	&	0.0	&&	54.2	&	10.4	&	0.0	\\
6fri26    	&&	13	&	50.0	&	0.0	&&	28.7	&	3	&	0.0	&&	50.0	&	20.3	&	0.0	\\
6bayg29   	&&	12	&	41.4	&	0.0	&&	37.9	&	5	&	0.0	&&	41.4	&	33.6	&	0.0	\\
9dantzig42	&&	6	&	14.3	&	0.0	&&	36.2	&	0	&	0.0	&&	14.3	&	24.9	&	0.0	\\
10att48   	&&	15	&	31.3	&	0.0	&&	41.5	&	7	&	0.0	&&	31.3	&	25.3	&	0.0	\\
10gr48    	&&	18	&	37.5	&	0.0	&&	27.0	&	4	&	0.0	&&	37.5	&	25.5	&	0.0	\\
10hk48    	&&	6	&	12.5	&	0.0	&&	34.2	&	3	&	0.0	&&	12.5	&	32.3	&	0.0	\\
11berlin52	&&	15	&	28.8	&	0.0	&&	36.1	&	1	&	0.0	&&	28.8	&	35.0	&	0.0	\\
11eil51   	&&	9	&	17.6	&	0.0	&&	32.6	&	3	&	0.0	&&	17.6	&	28.8	&	0.0	\\
12brazil58	&&	14	&	24.1	&	0.0	&&	24.5	&	3	&	0.0	&&	24.1	&	29.0	&	0.0	\\
14st70    	&&	12	&	17.1	&	0.0	&&	36.5	&	3	&	0.0	&&	17.1	&	24.6	&	0.0	\\
16eil76   	&&	12	&	15.8	&	0.0	&&	28.8	&	2	&	0.0	&&	15.8	&	28.6	&	0.0	\\
16pr76    	&&	2	&	2.6	&	0.0	&&	29.0	&	1	&	0.0	&&	2.6	&	29.7	&	0.0	\\
20gr96    	&&	13	&	13.5	&	0.0	&&	25.8	&	3	&	0.0	&&	13.5	&	20.6	&	0.0	\\
20rat99   	&&	11	&	11.1	&	0.0	&&	23.7	&	3	&	0.0	&&	11.1	&	23.2	&	0.0	\\
20kroA100 	&&	16	&	16.0	&	0.0	&&	20.9	&	2	&	0.0	&&	16.0	&	18.8	&	0.0	\\
20kroB100 	&&	8	&	8.0	&	0.0	&&	28.1	&	2	&	0.0	&&	8.0	&	25.0	&	0.0	\\
20kroC100 	&&	19	&	19.0	&	0.0	&&	27.2	&	2	&	0.0	&&	19.0	&	24.2	&	0.0	\\
20kroD100 	&&	19	&	19.0	&	0.0	&&	27.9	&	2	&	0.0	&&	19.0	&	19.8	&	0.0	\\
20kroE100 	&&	21	&	21.0	&	0.0	&&	26.4	&	1	&	0.0	&&	21.0	&	20.2	&	0.0	\\
20rd100   	&&	11	&	11.0	&	0.0	&&	32.1	&	2	&	0.0	&&	11.0	&	28.8	&	0.0	\\
21eil101  	&&	14	&	13.9	&	0.0	&&	35.5	&	1	&	0.0	&&	13.9	&	31.5	&	0.0	\\
21lin105  	&&	9	&	8.6	&	0.0	&&	35.4	&	3	&	0.0	&&	8.6	&	32.4	&	0.0	\\
22pr107   	&&	9	&	8.4	&	0.0	&&	35.6	&	0	&	0.0	&&	8.4	&	35.9	&	0.0	\\
24gr120   	&&	15	&	12.5	&	0.0	&&	28.4	&	4	&	0.0	&&	12.5	&	29.6	&	0.0	\\
25pr124   	&&	17	&	13.7	&	0.0	&&	32.5	&	3	&	0.0	&&	13.7	&	22.2	&	0.0	\\
26bier127 	&&	2	&	1.6	&	0.0	&&	21.5	&	1	&	0.0	&&	1.6	&	19.7	&	0.0	\\
26ch130   	&&	16	&	12.3	&	0.0	&&	25.9	&	3	&	0.0	&&	12.3	&	21.2	&	0.0	\\
28pr136   	&&	14	&	10.3	&	0.0	&&	22.4	&	1	&	0.0	&&	10.3	&	26.3	&	0.0	\\
28gr137   	&&	10	&	7.3	&	0.0	&&	19.9	&	1	&	0.0	&&	7.3	&	17.0	&	0.0	\\
29pr144   	&&	19	&	13.2	&	0.0	&&	33.2	&	2	&	0.0	&&	13.2	&	31.1	&	0.0	\\
30ch150   	&&	22	&	14.7	&	0.0	&&	19.9	&	2	&	0.0	&&	14.7	&	18.1	&	0.0	\\
30kroA150 	&&	20	&	13.3	&	0.0	&&	22.5	&	6	&	0.0	&&	13.3	&	19.5	&	0.0	\\
30kroB150 	&&	14	&	9.3	&	0.0	&&	23.8	&	2	&	0.0	&&	9.3	&	23.4	&	0.0	\\
31pr152   	&&	34	&	22.4	&	0.0	&&	37.5	&	7	&	0.0	&&	22.4	&	26.6	&	0.0	\\
32u159    	&&	33	&	20.8	&	0.0	&&	23.5	&	3	&	0.0	&&	20.8	&	15.1	&	0.0	\\
35si175   	&&	45	&	25.7	&	0.0	&&	27.4	&	5	&	0.0	&&	25.7	&	17.5	&	0.0	\\
36brg180  	&&	97	&	53.9	&	0.0	&&	57.9	&	51	&	0.0	&&	53.9	&	16.9	&	0.0	\\
39rat195  	&&	12	&	6.2	&	0.0	&&	22.2	&	1	&	0.0	&&	6.2	&	20.4	&	0.0	\\
40d198    	&&	7	&	3.5	&	0.0	&&	23.1	&	4	&	0.0	&&	3.5	&	24.2	&	0.0	\\
40kroA200 	&&	16	&	8.0	&	0.0	&&	20.3	&	2	&	0.0	&&	8.0	&	20.6	&	0.0	\\
40kroB200 	&&	7	&	3.5	&	0.0	&&	19.1	&	1	&	0.0	&&	3.5	&	18.5	&	0.0	\\
41gr202   	&&	4	&	2.0	&	0.0	&&	18.8	&	1	&	0.0	&&	2.0	&	18.5	&	0.0	\\
45ts225   	&&	40	&	17.8	&	0.0	&&	20.0	&	2	&	0.0	&&	17.8	&	11.2	&	0.0	\\
45tsp225  	&&	12	&	5.3	&	0.0	&&	20.5	&	2	&	0.0	&&	5.3	&	17.1	&	0.0	\\
46pr226   	&&	12	&	5.3	&	0.0	&&	29.6	&	1	&	0.0	&&	5.3	&	28.4	&	0.0	\\
46gr229   	&&	1	&	0.4	&	0.0	&&	22.0	&	0	&	0.0	&&	0.4	&	21.6	&	0.0	\\
53gil262  	&&	16	&	6.1	&	0.0	&&	21.8	&	3	&	0.0	&&	6.1	&	18.9	&	0.0	\\
53pr264   	&&	11	&	4.2	&	0.0	&&	21.5	&	1	&	0.0	&&	4.2	&	20.7	&	0.0	\\
56a280    	&&	20	&	7.1	&	0.0	&&	19.4	&	1	&	0.0	&&	7.1	&	16.1	&	0.0	\\
60pr299   	&&	15	&	5.0	&	0.0	&&	16.2	&	0	&	0.0	&&	5.0	&	14.7	&	0.0	\\
64lin318  	&&	13	&	4.1	&	0.0	&&	20.5	&	2	&	0.0	&&	4.1	&	20.8	&	0.0	\\
64linhp318	&&	13	&	4.1	&	0.0	&&	20.5	&	2	&	0.0	&&	4.1	&	20.8	&	0.0	\\
80rd400   	&&	11	&	2.8	&	0.0	&&	14.8	&	1	&	0.1	&&	2.8	&	13.0	&	0.0	\\
84fl417   	&&	43	&	10.3	&	0.0	&&	28.3	&	5	&	0.1	&&	10.3	&	22.7	&	0.1	\\
87gr431   	&&	0	&	0.0	&	0.0	&&	17.2	&	0	&	0.3	&&	0.0	&	17.2	&	0.3	\\
88pr439   	&&	10	&	2.3	&	0.0	&&	14.7	&	1	&	0.2	&&	2.3	&	15.0	&	0.1	\\
89pcb442  	&&	24	&	5.4	&	0.0	&&	11.9	&	0	&	0.1	&&	5.4	&	9.7	&	0.1	\\
99d493    	&&	4	&	0.8	&	0.0	&&	17.8	&	1	&	0.2	&&	0.8	&	19.4	&	0.2	\\
107att532 	&&	21	&	3.9	&	0.0	&&	20.5	&	2	&	0.3	&&	3.9	&	18.1	&	0.3	\\
107ali535 	&&	29	&	5.4	&	0.1	&&	16.6	&	2	&	0.5	&&	5.4	&	14.3	&	0.5	\\
107si535  	&&	96	&	17.9	&	0.0	&&	26.5	&	9	&	0.3	&&	17.9	&	17.9	&	0.1	\\
113pa561  	&&	147	&	26.2	&	0.1	&&	31.3	&	5	&	0.3	&&	26.2	&	22.6	&	0.1	\\
115u574   	&&	11	&	1.9	&	0.0	&&	14.4	&	1	&	0.2	&&	1.9	&	14.0	&	0.2	\\
115rat575 	&&	18	&	3.1	&	0.0	&&	11.2	&	2	&	0.2	&&	3.1	&	10.9	&	0.1	\\
131p654   	&&	88	&	13.5	&	0.1	&&	32.6	&	2	&	0.8	&&	13.5	&	28.2	&	0.5	\\
132d657   	&&	8	&	1.2	&	0.0	&&	10.8	&	0	&	0.3	&&	1.2	&	9.6	&	0.3	\\
134gr666  	&&	0	&	0.0	&	0.0	&&	11.6	&	0	&	1.0	&&	0.0	&	11.6	&	1.0	\\
145u724   	&&	34	&	4.7	&	0.1	&&	10.1	&	3	&	0.5	&&	4.7	&	8.8	&	0.4	\\
157rat783 	&&	25	&	3.2	&	0.0	&&	9.8	&	2	&	0.4	&&	3.2	&	8.4	&	0.3	\\
200dsj1000	&&	8	&	0.8	&	0.1	&&	9.6	&	1	&	2.4	&&	0.8	&	9.4	&	1.5	\\
201pr1002 	&&	20	&	2.0	&	0.1	&&	9.2	&	2	&	3.0	&&	2.0	&	8.7	&	1.6	\\
207si1032 	&&	85	&	8.2	&	0.2	&&	12.1	&	12	&	1.2	&&	8.2	&	10.2	&	0.9	\\
212u1060  	&&	36	&	3.4	&	0.1	&&	14.4	&	1	&	1.7	&&	3.4	&	11.2	&	2.0	\\
217vm1084 	&&	241	&	22.2	&	0.6	&&	24.0	&	8	&	2.3	&&	22.2	&	8.9	&	1.3	\\
235pcb1173	&&	11	&	0.9	&	0.1	&&	8.2	&	0	&	1.5	&&	0.9	&	8.2	&	1.3	\\
259d1291  	&&	48	&	3.7	&	0.2	&&	12.4	&	2	&	2.3	&&	3.7	&	9.8	&	1.7	\\
261rl1304 	&&	19	&	1.5	&	0.2	&&	7.9	&	2	&	2.6	&&	1.5	&	7.2	&	2.0	\\
265rl1323 	&&	23	&	1.7	&	0.2	&&	7.8	&	1	&	4.1	&&	1.7	&	7.0	&	2.9	\\
276nrw1379	&&	11	&	0.8	&	0.2	&&	7.4	&	1	&	3.7	&&	0.8	&	7.1	&	2.6	\\
280fl1400 	&&	23	&	1.6	&	0.9	&&	17.4	&	0	&	6.5	&&	1.6	&	17.5	&	5.3	\\
287u1432  	&&	33	&	2.3	&	0.2	&&	7.7	&	1	&	3.2	&&	2.3	&	6.6	&	2.6	\\
316fl1577 	&&	44	&	2.8	&	0.4	&&	10.3	&	2	&	5.0	&&	2.8	&	9.2	&	4.5	\\
331d1655  	&&	14	&	0.8	&	0.2	&&	6.7	&	1	&	3.7	&&	0.8	&	6.7	&	3.7	\\
350vm1748 	&&	285	&	16.3	&	2.5	&&	19.8	&	2	&	11.4	&&	16.3	&	11.0	&	5.5	\\
364u1817  	&&	5	&	0.3	&	0.1	&&	6.2	&	0	&	4.9	&&	0.3	&	5.8	&	4.5	\\
378rl1889 	&&	17	&	0.9	&	0.7	&&	7.3	&	3	&	10.9	&&	0.9	&	6.8	&	7.2	\\
421d2103  	&&	8	&	0.4	&	0.2	&&	6.7	&	1	&	2.9	&&	0.4	&	6.6	&	2.7	\\
431u2152  	&&	10	&	0.5	&	0.3	&&	5.2	&	0	&	7.8	&&	0.5	&	5.0	&	6.6	\\
464u2319  	&&	24	&	1.0	&	0.6	&&	3.9	&	0	&	10.3	&&	1.0	&	3.8	&	9.7	\\
479pr2392 	&&	33	&	1.4	&	0.9	&&	5.9	&	1	&	15.4	&&	1.4	&	5.3	&	13.4	\\
608pcb3038	&&	29	&	1.0	&	1.4	&&	4.7	&	1	&	45.4	&&	1.0	&	4.7	&	36.2	\\
759fl3795 	&&	21	&	0.6	&	4.9	&&	6.4	&	0	&	127.2	&&	0.6	&	6.5	&	94.5	\\
893fnl4461	&&	22	&	0.5	&	3.4	&&	3.1	&	0	&	80.2	&&	0.5	&	2.9	&	46.7	\\
1183rl5915  	&&	28	&	0.5	&	7.9	&&	2.4	&	2	&	258.1	&&	0.5	&	2.3	&	114.1	\\
1187rl5934  	&&	38	&	0.6	&	9.4	&&	3.0	&	2	&	308.3	&&	0.6	&	2.7	&	139.6	\\
1480pla7397 	&&	196	&	2.6	&	31.5	&&	4.6	&	1	&	2147.9	&&	2.6	&	3.6	&	1001.3	\\
2370rl11849 	&&	37	&	0.3	&	40.7\\
2702usa13509	&&	21	&	0.2	&	98.7\\
\bottomrule{}
\end{longtable}

The results of the experiments show that the preprocessing time for the Vertex Reduction is negligible for all the instances up to \texttt{212u1060}, i.e., for almost all \texttt{TSPLIB}-based GTSP instances used in the literature. The average percentage of detected redundant vertices for these instances is 14\%, and it is 11\% for all considered instances.  The experimental algorithm complexity is about $O(N^{2.4})$.

The Edge Reduction is more time-consuming than the Vertex Reduction.  The running time is negligible for all instances up to \texttt{115rat575}.  Note that in most of the GTSP literature, only instances with $N < 500$ are considered. The average per cent
of the detected redundant edges for these instances is about 27\%,
and it is 21\% for all instances in Table~\ref{tab:pure}. The
experimental algorithm's complexity is $O(N^{2.6})$.

\subsection{Algorithms Application Results}

Certainly, one can doubt the usefulness of our reduction algorithms since they may not necessarily decrease the running time of GTSP solvers.  Therefor, we tested the improvement of the running time of the following GTSP solvers:
\begin{enumerate}
    \item Exact algorithm (\Exact{}) based on a transformation of GTSP to TSP~\cite{Transformation}; the algorithm from~\cite{fischetti2002} was not available.  The algorithm that we use converts a GTSP instance with $N$ vertices to a TSP instance with $3 N$ vertices in polynomial time, solves the obtained TSP using the \textsf{Concorde} solver~\cite{Concorde}, and then converts the obtained TSP solution to GTSP solution also in polynomial time.

    \item Memetic algorithm from~\cite{RandomKeyGA} (\SD{}).  A memetic algorithm (MA) is a combination of a genetic algorithm with local search.

    \item MA from~\cite{GutinKarapetyanGA} (\GKK{}).

    \item MA from~\cite{GoldenGA} (\SG{}).

    \item A modified version of MA from~\cite{GutinKarapetyanGA2009}, the state-of-the-art GTSP memetic solver, (\GK{}).
\end{enumerate}


Each test was repeated ten times. The columns of the tables not described in Section \ref{PureResults} are as follows:
\begin{itemize}
    \item $T_0$ is the initial problem solution time.
    \item $B$ is the time benefit, i.e., $(T_0 - T_\text{pr}) / T_0$, where $T_\text{pr}$ is the preprocessed problem solution time; it includes preprocessing time as well.
\end{itemize}


\begin{table}[ht] \centering
\caption{Time benefit for \Exact{}.} 
\label{tab:exact}
\begin{tabular}{l r c r r c r r c r r r }
\toprule{}
&
    && \multicolumn{2}{c}{Vertices Red.}
    && \multicolumn{2}{c}{Edge Red.}
    && \multicolumn{3}{c}{Combined Reduction} \\

\cmidrule{4-5}  \cmidrule{7-8}  \cmidrule{10-12}

Instance & $T_0$, sec
    &&  $R_v$, \% & $B$, \%
    &&  $R_e$, \% & $B$, \%
    &&  $R_v$, \% & $R_e$, \% & $B$, \% \\
\cmidrule{1-12}
5gr21     	&	0.8	&&	38.1	&	40	&&	48.7	&	52	&&	38.0	&	45.0	&	56	\\
5ulysses22	&	1.7	&&	50.0	&	60	&&	44.3	&	48	&&	50.0	&	39.5	&	79	\\
5gr24     	&	0.2	&&	54.2	&	74	&&	33.1	&	53	&&	54.1	&	10.4	&	81	\\
6fri26    	&	0.9	&&	50.0	&	67	&&	28.7	&	18	&&	50.0	&	20.3	&	74	\\
6bayg29	&	6.0	&&	41.4	&	19	&&	0.0	&	59	&&	41.3	&	33.6	&	70	\\
10gr48    	&	16.1	&&	37.5	&	57	&&	27.0	&	2	&&	37.5	&	25.5	&	55	\\
10hk48    	&	52.7	&&	12.5	&	16	&&	34.2	&	6	&&	12.5	&	32.3	&	22	\\
11eil51	&	32.8	&&	17.6	&	37	&&	32.6	&	17	&&	17.6	&	28.8	&	42	\\
14st70 	&	150.4	&&	17.1	&	43	&&	36.5	&	17	&&	17.1	&	24.6	&	50	\\
\cmidrule{1-12}																	
Average	&	&&	35.4	&	45.9	&&	31.7	&	30.2	&&	35.3	&	28.9	&	58.8	\\
\bottomrule{}
\end{tabular}
\end{table}

\begin{table}[ht] \centering
\caption{Time benefit for \GKK{}.}
\label{tab:gkk}
\begin{tabular}{l r c r r c r r c r r r }
\toprule{}
&
    && \multicolumn{2}{c}{Vertices Red.}
    && \multicolumn{2}{c}{Edge Red.}
    && \multicolumn{3}{c}{Combined Reduction} \\

\cmidrule{4-5}  \cmidrule{7-8}  \cmidrule{10-12}

Instance & $T_0$, sec
    &&  $R_v$, \% & $B$, \%
    &&  $R_e$, \% & $B$, \%
    &&  $R_v$, \% & $R_e$, \% & $B$, \% \\
\cmidrule{1-12}
89pcb442 	&	60.7	&&	5.4	&	4	&&	11.9	&	17	&&	5.4	&	9.7	&	35	\\
99d493   	&	85.2	&&	0.8	&	14	&&	17.8	&	19	&&	0.8	&	19.4	&	29	\\
107att532	&	101.2	&&	3.9	&	9	&&	20.5	&	20	&&	3.9	&	18.1	&	20	\\
107ali535	&	99.3	&&	5.4	&	0	&&	16.6	&	47	&&	5.4	&	14.3	&	51	\\
107si535 	&	166.1	&&	17.9	&	12	&&	26.5	&	14	&&	17.9	&	17.9	&	41	\\
113pa561  	&	101.8	&&	26.2	&	15	&&	31.3	&	21	&&	26.2	&	22.6	&	47	\\
115u574  	&	103.6	&&	1.9	&	-3	&&	14.4	&	12	&&	1.9	&	14.0	&	28	\\
115rat575 	&	219.3	&&	3.1	&	38	&&	11.2	&	36	&&	3.1	&	10.9	&	45	\\
131p654   	&	165.4	&&	13.4	&	21	&&	32.6	&	12	&&	13.4	&	28.2	&	38	\\
132d657  	&	189.1	&&	1.2	&	10	&&	10.8	&	22	&&	1.2	&	9.6	&	24	\\
134gr666  	&	224.8	&&	0.0	&	26	&&	11.6	&	36	&&	0.0	&	11.6	&	57	\\
145u724  	&	232.9	&&	4.6	&	25	&&	10.1	&	29	&&	4.6	&	8.8	&	55	\\
157rat783 	&	392.7	&&	3.1	&	1	&&	9.8	&	16	&&	3.1	&	8.4	&	29	\\
200dsj1000	&	898	&&	0.8	&	6	&&	9.6	&	52	&&	0.8	&	9.4	&	51	\\
\cmidrule{1-12}																	
Average	&		&&	6.3	&	12.7	&&	16.8	&	25.2	&&	6.3	&	14.5	&	39.3	\\
\bottomrule{}
\end{tabular}
\end{table}

\begin{table}[ht] \centering
\caption{Time benefit for \SD{}.}
\label{tab:sd}
\begin{tabular}{l r c r r c r r c r r r }
\toprule{}
&
    && \multicolumn{2}{c}{Vertices Red.}
    && \multicolumn{2}{c}{Edge Red.}
    && \multicolumn{3}{c}{Combined Reduction} \\

\cmidrule{4-5}  \cmidrule{7-8}  \cmidrule{10-12}

Instance & $T_0$, sec
    &&  $R_v$, \% & $B$, \%
    &&  $R_e$, \% & $B$, \%
    &&  $R_v$, \% & $R_e$, \% & $B$, \% \\
\cmidrule{1-12}
157rat783 	&	23.6	&&	3.2	&	11	&&	9.8	&	5	&&	3.1	&	8.4	&	36	\\
200dsj1000	&	100.3	&&	0.8	&	47	&&	9.6	&	36	&&	0.8	&	9.4	&	42	\\
201pr1002	&	54.9	&&	1.9	&	12	&&	9.2	&	22	&&	1.9	&	8.7	&	43	\\
207si1032	&	21.3	&&	8.2	&	3	&&	12.1	&	-1	&&	8.2	&	10.2	&	24	\\
212u1060 	&	88.8	&&	3.3	&	8	&&	14.4	&	35	&&	3.3	&	11.2	&	42	\\
217vm1084 	&	78.1	&&	22.2	&	49	&&	24.0	&	-2	&&	22.2	&	8.9	&	57	\\
235pcb1173	&	107.9	&&	0.9	&	5	&&	8.2	&	30	&&	0.9	&	8.2	&	32	\\
259d1291	&	169.4	&&	3.7	&	9	&&	12.4	&	25	&&	3.7	&	9.8	&	26	\\
261rl1304 	&	140.4	&&	1.5	&	9	&&	7.9	&	47	&&	1.4	&	7.2	&	66	\\
265rl1323 	&	132.6	&&	1.8	&	20	&&	7.8	&	20	&&	1.7	&	7.0	&	32	\\
276nrw1379	&	111.5	&&	0.8	&	4	&&	7.4	&	22	&&	0.7	&	7.1	&	46	\\
\cmidrule{1-12}
Average	&		&&	4.4	&	16.1	&&	11.2	&	21.7	&&	4.4	&	8.7	&	40.5	\\
\bottomrule{}
\end{tabular}
\end{table}

\begin{table}[ht] \centering
\caption{Time benefit for \SG{}.}
\label{tab:sg}
\begin{tabular}{l r c r r}
\toprule{}
&
    && \multicolumn{2}{c}{Vertices Red.}\\

\cmidrule{4-5}

Instance & $T_0$, sec
    &&  $R_v$, \% & $B$, \%\\
\cmidrule{1-5}
84fl417.gtsp	&	4.5	&&	10.3	&	12	\\
87gr431.gtsp	&	8.3	&&	0.0	&	6	\\
88pr439.gtsp	&	10.2	&&	2.3	&	-3	\\
89pcb442.gtsp	&	11.5	&&	5.4	&	0	\\
99d493.gtsp	&	20.0	&&	0.8	&	7	\\
107att532.gtsp	&	25.1	&&	3.9	&	11	\\
107si535.gtsp	&	16.9	&&	17.9	&	34	\\
107ali535.gtsp	&	29.1	&&	5.4	&	20	\\
113pa561.gtsp	&	14.5	&&	26.2	&	31	\\
\cmidrule{1-5}
Average	&		&&	8.0	&	13	\\
\bottomrule{}
\end{tabular}
\end{table}

\begin{table}[ht] \centering
\caption{Time benefit for \GK{}.}
\label{tab:gk}
\begin{tabular}{l r c r r c r r c r r r }
\toprule{}
&
    && \multicolumn{2}{c}{Vertices Red.}
    && \multicolumn{2}{c}{Edge Red.}
    && \multicolumn{3}{c}{Combined Reduction} \\

\cmidrule{4-5}  \cmidrule{7-8}  \cmidrule{10-12}

Instance & $T_0$, sec
    &&  $R_v$, \% & $B$, \%
    &&  $R_e$, \% & $B$, \%
    &&  $R_v$, \% & $R_e$, \% & $B$, \% \\
\cmidrule{1-12}
89pcb442.gtsp	&	3.43	&&	5.4	&	16	&&	12.0	&	-2	&&	5.4	&	9.8	&	7	\\
99d493.gtsp	&	6.36	&&	0.8	&	2	&&	17.9	&	0	&&	0.8	&	19.4	&	2	\\
107att532.gtsp	&	5.96	&&	3.9	&	7	&&	20.6	&	10	&&	3.9	&	18.1	&	11	\\
107si535.gtsp	&	4.52	&&	17.9	&	14	&&	26.5	&	8	&&	17.9	&	18.0	&	15	\\
107ali535.gtsp	&	8.91	&&	5.4	&	17	&&	16.6	&	19	&&	5.4	&	14.3	&	25	\\
113pa561.gtsp	&	6.86	&&	26.2	&	20	&&	31.3	&	6	&&	26.2	&	22.6	&	23	\\
115u574.gtsp	&	7.43	&&	1.9	&	-2	&&	14.4	&	-6	&&	1.9	&	14.0	&	-1	\\
115rat575.gtsp	&	7.29	&&	3.1	&	0	&&	11.3	&	0	&&	3.1	&	10.9	&	2	\\
131p654.gtsp	&	5.47	&&	13.5	&	11	&&	32.7	&	2	&&	13.5	&	28.3	&	13	\\
\cmidrule{1-12}																	
	&		&&	8.7	&	9	&&	20.4	&	4	&&	8.7	&	17.3	&	11	\\
\bottomrule{}
\end{tabular}
\end{table}

The experiments show that the Vertex Reduction, the Edge Reduction and the Combined Reduction Technique significantly reduce the running time of the \Exact{}, \SD{} and \GKK{} solvers.  However, the Edge Reduction (and because of that the Combined Reduction Techique) is not that successful for \SG{} (Table~\ref{tab:sg}) and the original version of \GK{}\@.  That is because not every algorithm processes infinite edges well.  

Next we show that a solver can be adjusted to work better with preprocessed instances.  For this purpose we modified \GK{} as follows:
\begin{itemize}
\item The 2-opt heuristic~\cite{GutinKarapetyanGA2009} was extended with the cluster optimization.  For every iteration of 2-opt, where edges $v_1 v_2$ and $v_3 v_4$ are removed, instead of replacing them with $v_1 v_3$ and $v_2 v_4$ we replace them with $v'_1 v_3$ and $v'_2 v_4$, where $v'_1 \in cluster(v_1)$ and $v'_2 \in cluster(v_2)$ and $v'_1$ and $v'_2$ are selected to minimize the solution objective value.  (Here $cluster(v)$ is the cluster corresponding to the vertex $v$: $v \in cluster(v)$.)  Thereby, while the initial 2-opt heuristic could decline some good 2-opt if $w(v_1 v_3) = \infty$ or $w(v_2 v_4) = \infty$, the extended 2-opt will pass round the infinite edges.

\item Direct 2-opt heuristic~\cite{GutinKarapetyanGA2009} is excluded from the Local Search Procedure.

\item Every time before starting the Cluster Optimization~\cite{GutinKarapetyanGA2009} we remove all vertices that cannot be included in the solution, i.e., if a fragment of the solution corresponds to clusters $C_1$, $C_2$ and $C_3$ and there is no edge from $C_1$ to $v \in C_2$ or there is no edge from $v$ to $C_3$ then $v$ can be excluded for the current Cluster Optimization run.

\item Since the modified Local Search Procedure is more powerful than the previous one, we reduced the number of solutions in a generation and the termination condition is also changed (now $r = 0.2G + 0.03M + 8$ while previously $r = 0.2G + 0.05M + 10$ and $I_\text{cur} \ge \max(1.5 I_\text{max}, 0.025M + 2)$ instead of $I_\text{cur} \ge \max(1.5 I_\text{max}, 0.05M + 5)$, see~\cite{GutinKarapetyanGA2009}).
\end{itemize}

The modified algorithm does not reproduce exactly the results of the initial \GK{} heuristic; it gives a little bit better solution quality at the cost of slightly larger running times.  However, one can see (Table~\ref{tab:gk}) that all the Reduction Algorithms proposed in this paper influence the modified \GK{} algorithm positively.

Different reductions have different degree of success for different solvers.  The Edge Reduction is more efficient than the Vertex Reduction for \GKK{} and \SD{}; in other cases the Vertex Reduction is more successful.  For every solver except \SG{} the Combined Technique is preferred to separate reductions.

\bigskip

Preprocessing is called to reduce the solution time.  On the other hand, there is no guaranty that the outcome of the preprocessing will be noticeable.  Thus, it is important to ensure at least that the preprocessing time is significantly shorter than the solution time.

Five GTSP solvers are considered in this paper.  The first solver, \Exact{}, is an exact one and, thus, it is clear that its time complexity is larger than $\Theta(N^{2.6})$ (see Section~\ref{PureResults}) or even the upper bound $O(N^3)$.  The time complexities of the other four solvers were estimated experimentally, i.e., experiments were conducted for problems of different size obtained from \texttt{TSPLIB}~\cite{TSPLIB} and then an approximation for ``solution time''/``instance size'' dependence was found.  The experimental complexity of \SD{} is about $\Theta(N^3)$ and it is about $\Theta(N^{3.5})$ for  \GK{}, \SG{} and \GKK{}\@.  Table \ref{tab:estimation} demonstrates the quality of our estimate for \SD{} (here $T_\text{estimate}(N) = 6.3319 \cdot 10^{-8} \cdot N^3$).
\begin{table}[!ht] \center
    \caption{\SD{} work time estimation.}
    \label{tab:estimation}
    \begin{tabular}{l r r}
	\toprule
    Instance name
        & Real solution time, sec
        & Estimation, sec \\
    \cmidrule{1-3}
    45ts225     & $ 0.6 $ & $   0.72    $ \\
    45tsp225    & $ 0.5 $ & $   0.72    $ \\
    46pr226     & $ 0.7 $ & $   0.73    $ \\
    46gr229     & $ 0.8 $ & $   0.76    $ \\
    53gil262    & $ 0.9 $ & $   1.14    $ \\
    53pr264     & $ 1.2 $ & $   1.17    $ \\
    60pr299     & $ 1.3 $ & $   1.69    $ \\
    64lin318    & $ 1.8 $ & $   2.04    $ \\
    64linhp318  & $ 1.6 $ & $   2.04    $ \\
    80rd400     & $ 3.5 $ & $   4.05    $ \\
    84fl417     & $ 3.5 $ & $   4.59    $ \\
    87gr431     & $ 3.7 $ & $   5.07    $ \\
    88pr439     & $ 4.7 $ & $   5.36    $ \\
    89pcb442    & $ 5.5 $ & $   5.47    $ \\
    107si535    & $ 5.7 $ & $   9.70 $ \\
    113pa561    & $ 6.7 $ & $   11.18   $ \\
    115u574     & $ 13.2    $ & $   11.97   $ \\
    115rat575   & $ 11.1    $ & $   12.04   $ \\
    131p654     & $ 10.2    $ & $   17.71   $ \\
    134gr666    & $ 14.0  $ & $   18.70    $ \\
    145u724     & $ 27.9    $ & $   24.03   $ \\
    157rat783   & $ 23.6    $ & $   30.40    $ \\
    200dsj1000  & $ 100.3   $ & $   63.32   $ \\
    201pr1002   & $ 54.9    $ & $   63.70    $ \\
    207si1032   & $ 21.3    $ & $   69.59   $ \\
    212u1060    & $ 88.8    $ & $   75.41   $ \\
    217vm1084   & $ 78.1    $ & $   80.65   $ \\
    235pcb1173  & $ 107.9   $ & $   102.19  $ \\
    259d1291    & $ 169.4   $ & $   136.24  $ \\
    261rl1304   & $ 140.4   $ & $   140.40   $ \\
    265rl1323   & $ 132.6   $ & $   146.63  $ \\
    276nrw1379  & $ 111.5   $ & $   166.05  $ \\
	\bottomrule
    \end{tabular}
\end{table}

Having the solvers time complexities, we can conclude that the preprocessing time is significantly smaller than the solution time for arbitrary large instances as the experimental complexity of preprocessing is smaller than the complexity of even the fastest of the considered solvers.  

\section{Conclusion}

The GTSP reduction techniques allow one to significantly decrease the problem complexity at a very low cost. Experiments show that the Combined Reduction is often the most powerful among the presented algorithms and takes even less time than the single Edge Reduction.  While the Vertex Reduction yields very natural problems and is successful with every considered solver, the Edge Reduction changes some edge weights to infinity values and, thus, not every solver benefits from it.  However, in this paper, it is shown that a solver can be modified to process such problems well.

In this paper we consider the symmetric case only, i.e., $\dist(x, y) = \dist(y, x)$ for every pair of vertices $x$ and $y$.  Other vertex and edge reduction algorithms that can be immediately derived from Definitions~\ref{definition:excess} and \ref{definition:excessedge} exist for the asymmetric case, and their time complexity is $O(N^3)$.  Recall that $N$ is the total number of problem vertices.

\bigskip

\noindent{\bf Acknowledgement} We would like to thank Larry Snyder and John Silberholz for kindly providing the source codes of \SD{} and \SG{}, respectively.

\end{document}